# Compact Multi-Service Antenna for Sensing and Communication Using Reconfigurable Complementary Spiral Resonator

Ali Raza[1, 2, 3], Rasool Keshavarz[1], Eryk Dutkiewicz[1], Negin Shariati[1, 2]

*Abstract*— In this paper, a compact multi-service antenna (MSA) is presented for sensing and communication using a reconfigurable complementary spiral resonator. A three turns complementary spiral resonator (3-CSR) is inserted in the ground plane of a modified patch antenna to create a miniaturized structure. Two Positive-Intrinsic-Negative (PIN) diodes (D1, D2) are also integrated with the 3-CSR to achieve frequency reconfiguration. The proposed structure operates in three different modes i.e., dual-band joint communication and sensing antenna (JCASA), dual-band antenna, and single-band antenna. The required mode can be selected by changing the state of the PIN diodes. In mode-1, the first band (0.95–0.97 GHz) of the antenna is dedicated to sensing by using frequency domain reflectometry (FDR), while the second band (1.53–1.56 GHz) is allocated to communication. The sensing ability of the proposed structure is utilized to measure soil moisture using FDR. Based on the frequency shift, permittivity of the soil is observed to measure soil moisture. In mode-2 and mode-3, the structure operates as a standard dual and single band antenna, respectively, with a maximum gain of 1.5 dBi at 1.55 GHz. The proposed planar structure, with its simple geometry and a high sensitivity of 1.7 %, is a suitable candidate for precision farming. The proposed structure is versatile and capable of being utilized as a single or dual-band antenna and also measuring permittivity of materials within the range of 1–20. Hence, it is adaptable to a range of applications.

## I. INTRODUCTION

The usage of Wireless Sensor Networks (WSNs) for automatic measurement has rapidly increased across various applications. The Internet-of-Things (IoT) WSNs reached a count of 50 billion in 2020, with a projected yearly increase of 12% year [1, 2]. As a part of IoT networks, WSNs have been widely used in various applications such as soil moisture measurement, atmosphere monitoring, and animal tracking for the implementation of smart agriculture [3]. The deployment of sensors in outdoor locations enables remote monitoring of soil parameters, such as moisture and humidity, without the need of physical visits. Therefore, the use of WSNs can enhance productivity by reducing manual labor and enabling remote monitoring. This technology can also be applied in livestock management [4], where wireless sensors can track animal locations and transmit data to the base station.

Radio frequency identification (RFID) tags and microwave sensors have gained significant attention in recent years [5, 6]. The basic principle of RFID technology involves using electromagnetic waves in the Ultra-High Frequency (UHF) band (0.840–0.955 GHz) to sense and transmit information from a tag to a base station. The capability to identify, measure, and communicate is crucial for effective functioning of RFID sensors [7]. To perform these functions, power is required, which can significantly affect the lifespan, cost, sensing range, and complexity of an RFID system [8-11]. Various RFID antenna designs have been proposed in the literature for different applications [12-18]. Recently, a dual-band circularly polarized (CP) crossed dipole antenna has been proposed for RFID applications [12]. The antenna with the size of $0.3 \times 0.3\ \lambda_0^2$, operates in the frequency range of 0.77–1.06 GHz and 2.22–2.95 GHz, but it exhibits a large physical size. Another dual-band CP antenna is presented for UHF RFID tag and wireless local area network (WLAN) applications in [13]. The antenna size is $0.18 \times 0.18\ \lambda_0^2$ with a gain of –0.6 dBi in the RFID band and 1.2 dBi in the WLAN band, but the geometry is non-planar.

RFID tags can be broadly divided into two categories: chipped and chipless [19]. In chipped RFID, an Application Specific Integrated Circuit (ASIC) chip is integrated with the structure for object identification [14-16]. A chipped UHF RFID tag antenna is presented in [17]. The antenna, with the dimensions of $0.23 \times 0.07\ \lambda_0^2$, operates in a single band (913–925 MHz) to detect metallic objects using an ASIC chip. A chipped RFID tag and sensor are presented for fluid detection [18]. Both the tag and the sensor are separately designed and are connected using a circulator to implement the sensing. Another chipped RFID fluid sensor is presented to sense the constitutive parameters of fluid [20]. Fluid flows between a capacitive gap and affects the capacitance of the sensor, which is then used to categorize the fluid, but the design process is complex. Whereas, in chipless tags, an ASIC chip is not required and detection is done using resonators [21-23]. Different types of resonators have been used in the literature to engineer chipless tags including slot resonators [24], spiral resonators [21], QR code-based resonators [22], natural resonance [25], and split ring resonators (SRR) [26].

[1]A. Raza, R. Keshavarz and N. Shariati are with RF and Communication Technologies (RFCT) research laboratory, School of Electrical and Data Engineering, Faculty of Engineering and IT, University of Technology Sydney, Ultimo, NSW 2007, Australia. (e-mail: Ali.Raza-1@student.uts.edu.au).
[1]E. Dutkiewicz is with the School of Electrical and Data Engineering, Faculty of Engineering and IT, University of Technology Sydney, Ultimo, NSW 2007, Australia.
[2]A. Raza and N. Shariati are also with Food Agility CRC Ltd, 175 Pitt St, Sydney, NSW, Australia 2000.
[3]A. Raza is a lecturer at the University of Engineering and Technology (UET) Lahore, Pakistan and is currently pursuing his full-time PhD while on study leave.

RFID microwave sensors have also been used to monitor soil moisture in smart agriculture by measuring soil permittivity. The amount of moisture in the soil is commonly referred to as Volumetric Water Content (VWC), and a higher value of VWC in the soil corresponds to a larger value of permittivity. The penetration of the RF signal into the soil depends on the frequency and dielectric properties of the soil [5]. The penetration of the signal can be increased by reducing the operating frequency of the sensor. This allows a single sensor to cover a larger volume of soil, thereby reducing the total number of sensors required in a wireless sensor network. However, lower frequencies result in bulkier structures that can be challenging to implement. Various types of sensors have been proposed in the literature for sensing, including capacitive sensors [27-30], frequency domain reflectometry (FDR) sensors [31-40], and time domain reflectometry sensors [41]. In capacitive sensors, an excitation signal is applied to the sensor and the capacitance value is measured based on the charging/discharging of the capacitor through the resistance. The capacitance value is then used to calculate the VWC in the soil, but the resistance value is sensitive to temperature variations, which can result in false measurements. A soil moisture sensor based on metamaterial absorber is presented in [32]. The absorption of the filter varies with different VWC levels in the soil at an operating frequency of 625 MHz. However, the sensor exhibits low sensitivity. Another soil moisture sensor based on FDR is presented in [33]. A combination of spiral resonators and complementary spiral resonators is used to measure soil moisture. However, the resonance frequency of the sensor is 4 GHz, which implies low penetration capacity of the signal, also the sensor is very sensitive to different volume under test (VUT) of the soil. Sensors for permittivity measurement based on complementary split ring resonator (CSRR) and complementary curved ring resonator (CCRR) are presented in [34, 39, 40]. These sensors operate at 2.67 GHz, 2.7 GHz, and 3.49 GHz, respectively, and demonstrate high sensitivity. However, the permittivity measurement depends on the thickness of the material under test (MUT), and distance between the sensor and MUT due to high operating frequency. CSRR based sensors are also presented for microfluid characterization [35-38]. The resonance frequency of the sensor changes with different microfluids, but the sensitivity is low. Recently, a microwave sensor and an antenna are integrated in a dual-function structure for sensing and communication [42]. The sensor operates at a higher frequency of 4.7 GHz, while antenna operates at 2.45 GHz. The sensor comprises of a frequency-selective filter to characterize different substrates, but the sensor lacks sharp resonances, which is necessary for accurate measurement, and it exhibits high frequency and low sensitivity.

The proposed compact multi-service antenna (MSA) operates in three modes: dual-band joint communication and sensing antenna (JCASA), dual-band antenna, and single-band antenna. A three turns complementary spiral resonator (3-CSR) is used with a modified patch to achieve miniaturization [43, 44]. Two Positive-Intrinsic-Negative (PIN) diodes (D1 and D2) are integrated with the 3-CSR structure and three different ON/OFF configurations have been used, i.e., 00, 10, and 11, with '0' representing the OFF state and '1' representing the ON state of the diode. The required mode can be selected by changing the configuration of diodes. For '00' case, the structure operates as dual-band JCASA and has the ability to sense and communicate. This mode is used to measure soil moisture by using FDR in the first band, while the second band is used for communication. The proposed MSA possesses an adaptive nature and can be used to measure the permittivity of any MUT within the range of 1–20. Due to its sensing and communication ability, the MSA is a good candidate for soil moisture measurement in precision farming. The proposed MSA is also suitable for standard single and dual band antenna applications.

Major contributions of this paper are summarized as follows:

- The proposed structure features a compact design with simple planar geometry, yet the resonance frequency of the unloaded sensor is 960 MHz. This characteristic makes it suitable for covering a large VUT of soil. Additionally, the structure includes a communication unit in mode-1 to transfer information to the base station.
- Besides joint sensing and communication, the integration of 3-CSR and PIN diodes results in a multi-service structure that can function as a standard single/dual band antenna.
- The proposed structure is adaptive and suitable to measure the permittivity of any MUT within the range of 1–20.
- The design procedure along with the equivalent model of MSA is presented as a design guide for future work in joint communication and sensing systems.

The organization of this paper is as follows: Design guide, theoretical analysis, and working principle of JCASA are provided in Section II. Simulation and measurement results are presented in Section III. Finally, the conclusion is provided in Section IV.

II. MULTI-SERVICE ANTENNA DESIGN, THEORY AND METHODOLOGY

The design, theory, and methodology of the proposed MSA are discussed in the following subsections:

A. *Multi-service Antenna Design*

The antenna is designed on Rogers RO4003C substrate ($\varepsilon_r = 3.55$, $tan\delta = 0.0027$, $h = 1.6$ mm) with dimensions of $50 \times 50$ mm$^2$. Initially, a conventional patch antenna was designed and simulated at 3.2 GHz in CST MWS 2019 as shown in Fig. 1. Fig. 1(a) represents the top side and Fig. 1(b) represents the bottom side of the patch antenna with the inset feed. To achieve dual resonance, two additional patches and two slots are inserted as shown in Fig. 1(c).

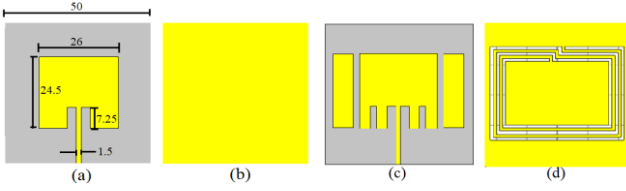

Fig. 1. Design procedure of the proposed antenna, (a) top view of the conventional patch, (b) bottom view of the conventional patch, (c) top view of the modified patch, (d) bottom view of the modified patch

This modification generated two resonances at 3.2 GHz and 3.4 GHz. In order to reduce the antenna size, a 3-CSR is inserted in the ground plane of the modified patch as shown in Fig. 1(d). The reflection coefficients of the antenna at different stages are shown in Fig. 2.

The resonance frequency of a resonator depends on the equivalent values of inductance ($L$) and capacitance ($C$) and by changing these values, the resonance frequency can be switched. To change the resonance frequency of the 3-CSR structure, two PIN diodes (D1 and D2) are integrated with the 3-CSR to realize a three turns reconfigurable complementary spiral resonator (3-RCSR). The optimal position for the PIN diodes is selected based on a parametric analysis of the frequency response, which is conducted by placing the diodes at various positions. The integration of the PIN diodes results in the achievement of another operating band at 0.96 GHz, which further reduces the size of the antenna. The diode model Skyworks SMP1322 is utilized for the design. In Section III, the simulation and measurement results of the MSA for different diode states are discussed.

The geometry of the proposed antenna is shown in Fig. 3, where Fig. 3(a) represents the modified patch and Fig. 3(b) represents the 3-RCSR. The diode is modeled as a lumped circuit in ON and OFF states. Different antenna applications are achieved by varying the state of the diodes i.e., 00, 10, and 11. DC blockers are introduced to prevent the shorting of the positive and negative pins of the Direct Current (DC) voltage as shown in Fig. 3. A biasing network is also designed for the diodes at the top side of the same substrate which is connected to the diodes using vias. The biasing circuit for each diode consists of two RF chokes ($L_1$ and $L_2$) and two resistors ($R_1$ and $R_2$).

*B. MSA Theory*

The equivalent circuit (EC) of a 3-CSR consists of an LC circuit with two series inductors ($L_0$) and a capacitor ($C_c$) as shown in Fig. 4(a) [45]. The ECs of the PIN diode are shown in Fig. 4(b) and Fig. 4(c). The resonance frequency can be calculated using (1), where $L_C$ is the equivalent inductance. The value of inductance ($L_0$) depends on the line impedance ($Z_0$), effective permittivity ($\varepsilon_{re}$), speed of light ($c$), and line length ($l$) as represented by (2) [46]. The values of $Z_0$ and $\varepsilon_{re}$ can be calculated using (3) and (4) [47]. The lengths ($l$) of the smallest and largest turns are 125 mm and 140 mm, respectively. The width ($W$) of a single turn is 1 mm and thickness of the substrate ($h$) is 1.6 mm.

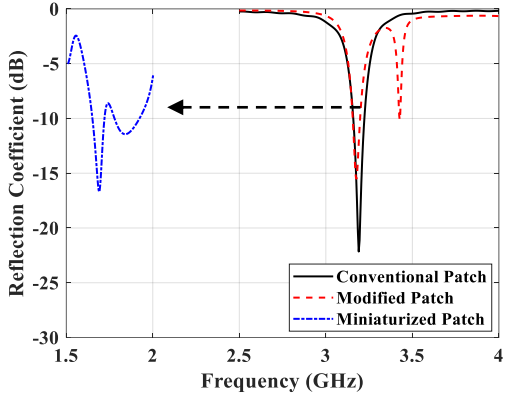

Fig. 2. Simulated reflection coefficients of the patch antenna

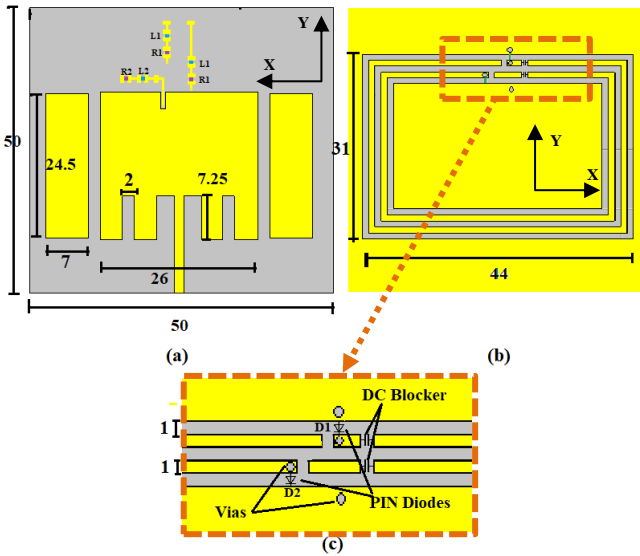

Fig. 3. Geometry of the proposed antenna, (a) top view, (b) bottom view, (c) magnified view (all dimensions are in mm)

$$f = \frac{1}{2\pi\sqrt{L_c C_c}} \quad (1)$$

$$L_0 = \frac{Z_0 \sqrt{\varepsilon_{re}}}{c} l \quad (2)$$

The integration of the PIN diodes with the 3-CSR modifies the behavior of the LC circuit, resulting in a dual-band (0.96/1.55 GHz) structure. Modified EC of the 3-RCSR is shown in Fig. 5 and the diodes can be modeled as lumped components as shown in Fig. 4. Different states of the diodes modify the values of inductance and capacitance and change the resonance frequency according to (1).

$W < h$:

$$\varepsilon_{re} = \frac{\varepsilon_r + 1}{2} + \frac{\varepsilon_r + 1}{2}\left[\frac{1}{\sqrt{1 + 12\left(\frac{h}{W}\right)}} + 0.04\left(1 - \frac{W}{h}\right)^2\right] \quad (3.a)$$

$$Z_0 = \frac{60}{\varepsilon_{re}} \log_2\left[8\frac{h}{W} + 0.25\frac{W}{h}\right] \quad (3.b)$$

$W > h$:

$$\varepsilon_{re} = \frac{\varepsilon_r + 1}{2} + \frac{\varepsilon_r + 1}{2}\left[\frac{1}{\sqrt{1 + 12\left(\frac{h}{W}\right)}}\right] \quad (4.a)$$

$$Z_0 = \frac{120\pi}{\varepsilon_{re}\left[\frac{W}{h} + 1.393 + \frac{2}{3}\log_2\left(\frac{W}{h} + 1.444\right)\right]} \quad (4.b)$$

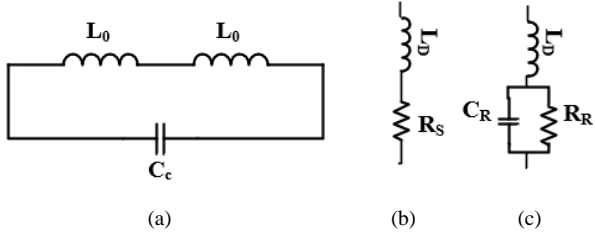

Fig. 4. Equivalent circuits, (a) 3-CSR, (b) PIN diode in ON State, and (c) PIN Diode in OFF State

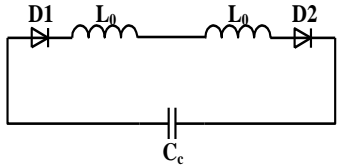

Fig. 5. Equivalent circuit of the 3-RCSR

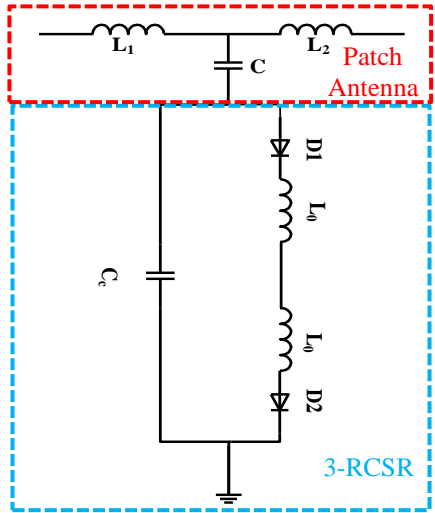

Fig. 6. Equivalent circuit of the proposed JCASA

Patch antenna at the top side of the substrate can be considered as a right-handed transmission line of physical length, consisting of series inductances and parallel capacitances. A complete circuit of the patch antenna with 3-RCSR is shown in Fig. 6, where the inductance of the patch can be calculated using (2).

### C. Working Principle of the JCASA

In RFID tags, a radio frequency (RF) reader sends a continuous wave (CW) interrogating signal for a short period of time. This interrogating signal is captured by the tag which encodes the signal and sends it back to the reader.

The proposed MSA structure has a frequency tunable feature, and antenna mode can be changed by switching the state of the PIN diodes. A specified state of the diodes can be used to select a required mode of the antenna. In the JCASA mode, the proposed structure is used for sensing as well as communication. A high-level block diagram of the proposed system is shown in Fig. 7. A CW signal is sent from an oscillator to the JCASA using a 3-port circulator, and the reflected signal is measured using a power detector. The real value of permittivity is measured using FDR. For the experimental purpose, the oscillator and power detector are replaced by a vector network analyzer as shown in Fig. 7(b). The frequency shift in the first band is analyzed to measure the permittivity [33, 34], while the second band is used to transfer the information to the base station. In general, the proposed JCASA can sense the permittivity of any object from 1 to 20.

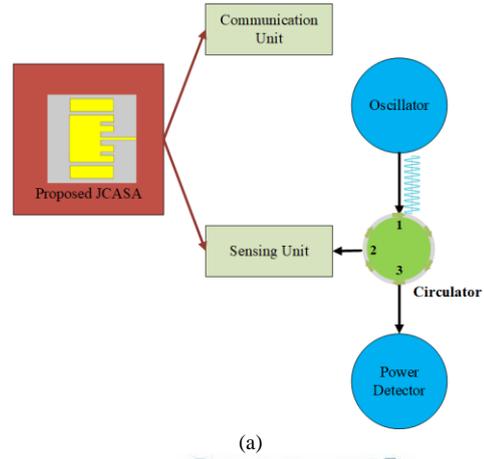

(a)

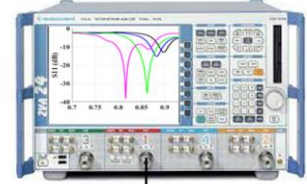

(b)

Fig. 7. Working principle of the proposed JCASA, (a) practical setup, (b) experimental lab setup

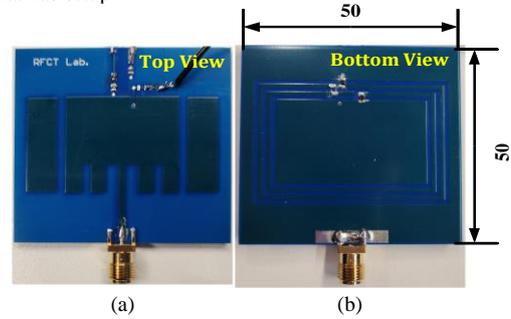

Fig. 8. Fabricated prototype of the antenna on Rogers substrate ($\varepsilon_r = 3.55$, $tan\delta = 0.0027, h = 1.6$ mm) with dimensions of $50 \times 50$ mm$^2$

### III. RESULTS AND DISCUSSION

To demonstrate the performance of the MSA, the proposed antenna is simulated, designed, and optimized using CST MWS 2019. The MSA is fabricated on a Rogers RO4003C substrate, and the fabricated prototype is shown in Fig. 8.

### A. Simulated and Measured MSA Results for Different States

The proposed MSA can be used as a JCASA, dual-band antenna, and single-band antenna. Simulated and measured reflection coefficients of the antenna for different diode states are shown in Fig. 9 and Fig. 10, respectively. Both

> REPLACE THIS LINE WITH YOUR PAPER IDENTIFICATION NUMBER (DOUBLE-CLICK HERE TO EDIT) <    4

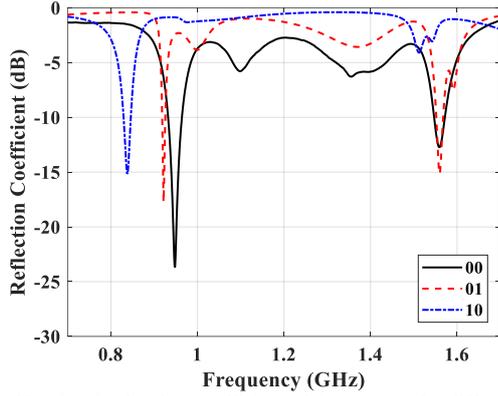
Fig. 9. Simulated reflection coefficient of the antenna for different states

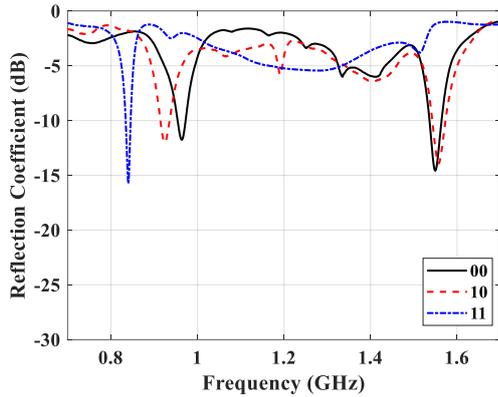
Fig. 10. Measured reflection coefficient of the antenna for different states

TABLE I
MEASURED BANDWIDTHS OF THE PROPOSED MSA FOR DIFFERENT STATES

| Mode | Diodes State (D2, D1) | Band 1 (GHz) | Band 2 (GHz) | Application |
|---|---|---|---|---|
| 1 | 00 | 0.95–0.97 | 1.53–1.56 | Dual-band JCASA |
| 2 | 10 | 0.91–0.94 | 1.54–1.57 | Dual-band antenna |
| 3 | 11 | 0.83–0.85 | – | Single-band antenna |

simulation and measurement results represent similar behavior, indicating the validity of the structure. Measured impedance bandwidths of the MSA are summarized in Table I against each state. The proposed MSA operates in three modes by switching the state of the PIN diodes. For '00' case (mode-1), the antenna operates as dual-band JCASA, where the first band (0.95–0.97 GHz) is used for sensing to measure the permittivity (1–20) and the second band (1.53–1.56 GHz) is allocated to communication. For '10' and '11', the proposed antenna operates in dual-band mode (0.91–0.94 GHz, 1.54–1.57 GHz) and single-band mode (0.83–0.85 GHz), respectively.

To validate the communication unit, two–dimensional (2D) gain patterns of the MSA are measured in *xz* and *yz* planes. Simulated and measured 2D patterns at 0.93 GHz and 1.55 GHz are shown in Fig. 11. The proposed antenna shows an omnidirectional radiation pattern and maximum gain of the unloaded antenna is 1.5 dBi at 1.55 GHz. The resonance frequency in the communication band changes from 1.54 to 1.35 GHz with the VWC of the soil. The radiation pattern of the antenna remains omnidirectional at all resonances, but a higher VWC reduces the antenna gain.

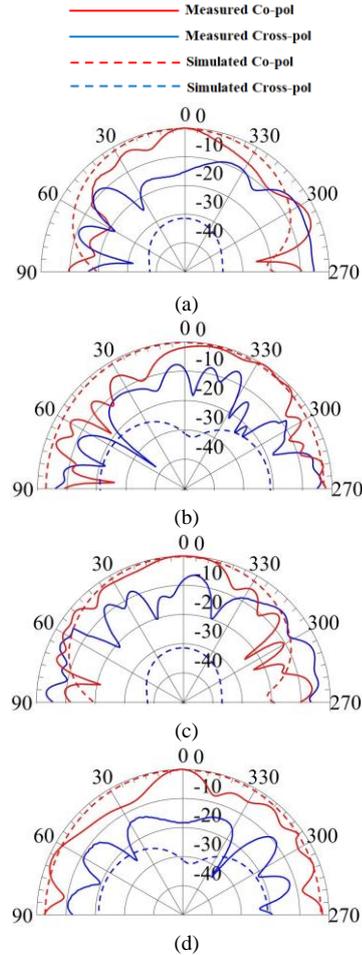
Fig. 11. Simulated and measured 2D Gain patterns of the proposed antenna, (a) *yz* at 0.93 GHz, (b) *xz* at 0.93 GHz, (c) *yz* at 1.55 GHz, and (d) *xz* at 1.55 GHz

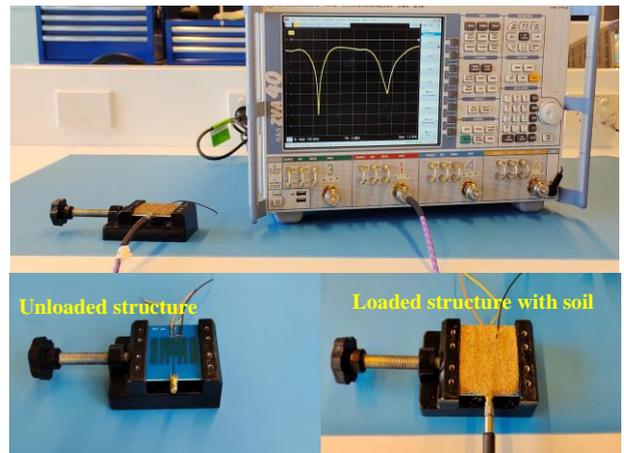
Fig. 12. Measurement setup for the sensing unit with unloaded and loaded structure

### B. Sensing Unit Results and Analysis

To verify the performance of the sensing unit in mode-1 for '00' case, an experimental setup shown in Fig. 12 was used to examine the fabricated prototype. The frequency responses of the structure are measured using a calibrated 4-port vector network analyzer (VNA-ZVA40) with different VWC levels in the soil. The VNA is calibrated using an SOLT (short, open, load and through) standard before executing the frequency responses. The proposed structure is placed in a non-metallic holder and the sensor depth is adjusted to 3 mm from the top of the holder to put soil on the structure. A sweep of RF signals is sent to the sensor and the reflected signal is measured in terms of S11 to estimate the permittivity of the MUT. The permittivity of the soil increases with the VWC and the value of permittivity ($\varepsilon'_r$) changes from 3.7 to 19 with a VWC range of 0 to 30 %, as shown in Table II. Fig. 13 shows the simulated and measured results for the '00' case. As the water content in the soil increases, the permittivity of the soil also increases, resulting a significant leftward shift in the resonance frequency. Hence, the frequency responses from 0.93 GHz to 0.83 GHz reflects the change in permittivity from 3.7 (VWC=0 %) to 19 (VWC=30 %). To assess the accuracy of the sensor, the measurement is repeated four times and a maximum variation of 10 MHz is observed in the frequency responses. A linear relationship between $\varepsilon'_r$ and frequency shift ($\Delta f$) for different tests is shown in Fig. 14. The frequency shift ($\Delta f$) is calculated using (5),

$$\Delta f = f_m - f_u \quad (5)$$

where $f_u$ is the resonance frequency of unloaded structure and $f_m$ is the resonance frequency of the structure with material under test.

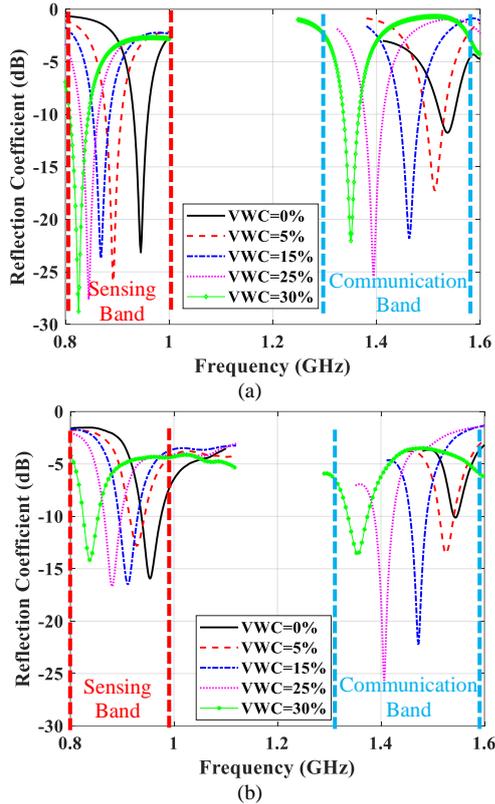

Fig. 13. Frequency response of MSA in mode-1 (JCASA) for different values of VWC, (a) simulated, (b) measured

TABLE II
REAL PERMITTIVITY OF SOIL FOR DIFFERENT VWC [48]

| VWC (%) | 0 | 5 | 10 | 15 | 20 | 25 | 30 |
|---|---|---|---|---|---|---|---|
| $\varepsilon'_r$ | 3.7 | 4.4 | 5.8 | 8.4 | 12 | 15.8 | 19 |

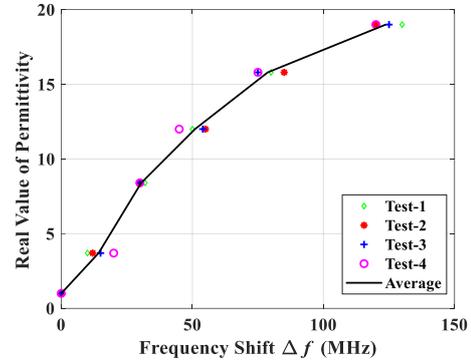

Fig. 14. Real value of permittivity ($\varepsilon'_r$) vs frequency shift ($\Delta f$)

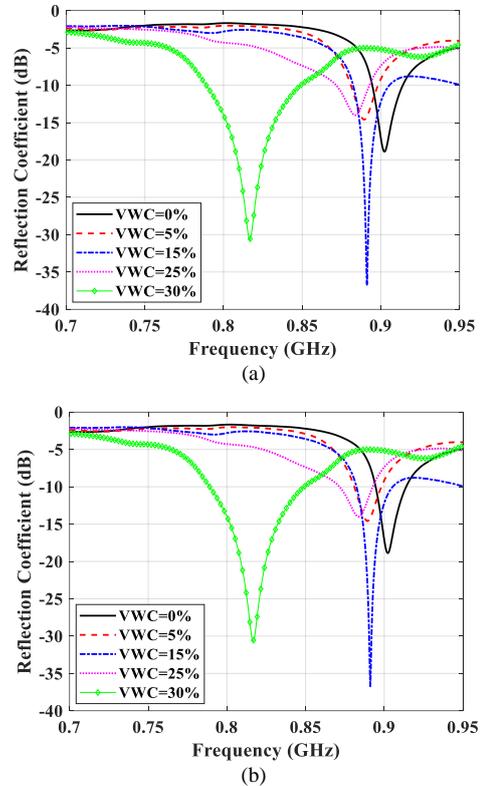

Fig. 15. Measured frequency response of the MSA for different values of VWC, (a) mode-2, (b) mode-3

The sensing ability of the MSA is also examined in mode-2 and mode-3 to validate the simulation results. Measured frequency responses of the antenna for '10' (mode-2) and '11' (mode-3) are shown in Fig. 15. In modes 3 and 4, the frequency shift does not follow a significant pattern with different values of VWC as evidenced by the results presented in Fig. 15. Hence, the proposed MSA can be switched to JCASA, dual-band antenna, and single-band antenna depending on the state of diodes.

To further evaluate the performance of the sensing unit, the sensitivity of the proposed structure is calculated using (6),

$$Sensitivity = \left|\frac{f_1 - f_2}{f_u(\varepsilon_{r1} - \varepsilon_{r2})}\right| \times 100 \quad (6)$$

where $f_1$ is the current resonance frequency, $f_2$ is the updated resonance frequency due to a new material, $\varepsilon_{r1}$ is the relative permittivity at $f_1$, $\varepsilon_{r2}$ is the relative permittivity at $f_2$ and $f_u$ is the resonance frequency of unloaded structure. A comparison of the proposed antenna in mode-1 (sensing) with those reported in the literature is summarized in Table III. In comparison to previously reported sensors [32-39, 42], the proposed structure boasts a compact design with good sensitivity. Furthermore, the proposed structure exhibits a dual-mode behavior, capable of functioning as a JCASA and a standard single/dual band antenna.

TABLE III
COMPARISON OF THE PROPOSED STRUCTURE WITH REPORTED SENSORS

| Ref. | Size at $f_u$ ($\lambda_0^2$) | $f_u$ (GHz) | Number of Bands | Application | Measurement Technique | Sensitivity at max ($\varepsilon_r$) (%) | Max Measured Permittivity |
|---|---|---|---|---|---|---|---|
| This Work | 0.158×0.158 | 0.95 | 2 | Sensing + Communication | 3-RCSR | 1.7 | 16.7 |
| [32] | 0.425×0.425 | 0.625 | 1 | Sensing | Metamaterial Absorber | 0.097 | 19.1 |
| [33] | 0.67×0.13 | 4 | 2 | Sensing | SRR and CSRR | 0.9 | 16.7 |
| [34] | 0.35× – | 2.67 | 3 | Sensing | CSRR | 1.6 | 9.2 |
| [35] | – | 2.4 | 1 | Sensing | CSRR | 0.19 | 79.5 |
| [36] | 0.32×0.2 | 2.45 | 1 | Sensing | M-CSRR | 0.2 | 70 |
| [37] | 0.198×0.198 | 2.38 | 1 | Sensing | EBG Resonator | 0.224 | 70 |
| [38] | 0.184×0.372 | 2.234 | 1 | Sensing | SRR | 0.04476 | 70 |
| [39] | 0.36× – | 2.7 | 1 | Sensing | CSRR | 1.7 | 10.2 |
| [42] | 0.42×0.44 | 4.7 | 2 | Sensing+ Communication | Frequency Selective Multipath Filter | 0.214 | 26 |

IV. CONCLUSION

In this paper, a compact multi–service antenna (MSA) is presented for sensing and wireless communication using a reconfigurable complementary spiral resonator. The antenna consists of a modified patch and a three turns complementary spiral resonator (3-CSR). Two PIN diodes are integrated with the 3-CSR to realize an MSA. The proposed antenna can operate in three modes: dual-band joint communication and sensing antenna (JCASA), dual-band antenna, and single-band antenna. In the JCASA mode, the first band (0.95–0.97 GHz) is used for sensing to precisely measure the permittivity while the second band (1.53–1.56 GHz) is allocated for communication. In mode-2 and 3, the proposed MSA operates as a dual band and single band antenna, respectively. The proposed structure is fabricated and measured to validate its performance and a favorable agreement being observed between the simulation and measurement results. Based on the experimental results, the proposed design is suitable for measuring soil moisture in precision farming, determining the permittivity of materials within the range of 1–20, and implementing single or dual-band antenna applications. Hence, the MSA is versatile and adaptable to a range of applications.

ACKNOWLEDGEMENT

This project was supported by funding from Food Agility CRC Ltd, funded under the Commonwealth Government CRC Program. The CRC Program supports industry-led collaborations between industry, researchers and the community.